\documentclass[a4paper,12pt]{article}
\usepackage[margin=0.9in]{geometry}
\usepackage[dvipsnames,svgnames]{xcolor}
\usepackage{graphicx}
\usepackage[edges]{forest}
\usepackage[frozencache,cachedir=.]{minted}
\usepackage[T1]{fontenc}
\usepackage[bitstream-charter]{mathdesign}
\title{A Subset of the CERN Virtual Machine File System: Fast Delivering of Complex Software Stacks for Supercomputing Resources}
\author{Alexandre F. Boyer$^{1,2}$, Christophe Haen$^2$, Federico Stagni$^2$, David R.C. Hill$^1$\\
\normalsize{\{alexandre.franck.boyer, christophe.haen, federico.stagni\}@cern.ch, david.hill@uca.fr}\\
\normalsize{1 Université Clermont Auvergne, Clermont Auvergne INP,}\\\normalsize{CNRS, Mines Saint-Etienne, LIMOS, 63000 Clermont-Ferrand, France}\\
\normalsize{2 European Organization for Nuclear Research, Meyrin, Switzerland}}
\date{}

\begin{document}

\maketitle              
\begin{abstract}
Delivering a reproducible environment along with complex and up-to-date software stacks on thousands of distributed and heterogeneous worker nodes is a critical task. The CernVM-File System (CVMFS) has been designed to help various communities to deploy software on worldwide distributed computing infrastructures by decoupling the software from the Operating System. However, the installation of this file system depends on a collaboration with system administrators of the remote resources and an HTTP connectivity to fetch dependencies from external sources. Supercomputers, which offer tremendous computing power, generally have more restrictive policies than grid sites and do not easily provide the mandatory conditions to exploit CVMFS. Different solutions have been developed to tackle the issue, but they are often specific to a scientific community and do not deal with the problem in its globality. In this paper, we provide a generic utility to assist any community in the installation of complex software dependencies on supercomputers with no external connectivity. The approach consists in capturing dependencies of applications of interests, building a subset of dependencies, testing it in a given environment, and deploying it to a remote computing resource. We experiment this proposal with a real use case by exporting Gauss - a Monte-Carlo simulation program from the LHCb experiment - on Mare Nostrum, one of the top supercomputers of the world. We provide steps to encapsulate the minimum required files and deliver a light and easy-to-update subset of CVMFS: 12.4 Gigabytes instead of 5.2 Terabytes for the whole LHCb repository.
\end{abstract}
\section{Introduction}
To study the constituents of matter and better understand the fundamental structure of the universe, HEP collaborations rely on complex software stacks and a worldwide distributed system to process a growing amount of data: the World Wide LHC Computing Grid (WLCG) \cite{WLCG}.
The infrastructure involves 170 computing centers, 1 million cores and 1 exabyte of storage spread around 42 countries.

Delivering a reproducible environment along with up-to-date software across thousands of heterogeneous computing resources is a major challenge: Buncic et al. designed CernVM and CVMFS (CernVM-File System) \cite{Buncic_2010} to tackle it by decoupling the software from the Operating System.

CernVM \cite{CVMFS} is a thin Virtual Software Appliance of about 150 Mb in its simplest form.
It supports a variety of hypervisors and container technologies and aims to provide a complete and portable user environment for developing and running HEP applications on any end-user computer and Grid Sites, independently of the underlying Operating Systems used by the targeted platforms.

CVMFS \cite{CVMFS} is a scalable and low-maintenance file system optimized for software distribution.
CVMFS is implemented as a POSIX read-only file system in user space.
Files and directories are hosted on standard web servers and mounted on the computing resources as a directory.
The file system performs aggressive file-level caching: both files and file metadata are cached on local disks as well as on shared proxy servers, allowing the file system to scale to a large number of clients \cite{Buncic_2010}.

This approach has been mainly adopted by the HEP community and is now getting users from various communities according to Arsuaga-Ríos et al. \cite{Arsuaga_R_os_2015}. In a few years, it has become the standard software distribution service on Grid Sites of WLCG.
Nevertheless, computing infrastructure and funding models are changing, and national science programs are consolidating computing resources and encourage using cloud systems as well as supercomputers, as Barreiro et al. explain \cite{Barreiro_2019}.
CVMFS developers have extended the features of the file system and have provided additional tools to support clouds \cite{Buncic_2011}\cite{Harutyunyan_2012} and supercomputers \cite{Blomer_2017}.

Supercomputers are highly heterogeneous architectures that pose higher integration challenges than traditional Grid Sites.
Many supercomputers do not allow a CVMFS client to be mounted on the worker nodes and/or do not provide external connectivity, which is critical to work with CVMFS.
CVMFS tools designed to interact with High-Performance Computing sites are aimed at administrators of scientific communities that would like to integrate their workflows on such machines: they ease some steps of the process but may require additional efforts on behalf of the administrators. 

In this study, we aim to automate the whole process and reduce these additional efforts by providing a utility able to extract, test and deploy parts of CVMFS on supercomputers not having outbound connectivity.
Section \ref{section:1} briefly introduces CVMFS and the ecosystem developed around it, in order to deal with supercomputers.
Section \ref{section:2} focuses on the design of the utility, the steps to extract software dependencies and to deploy them on a given supercomputer.
Finally, section \ref{section:3} presents a use case and the obtained results in detail.

\section{Context}\label{section:1}
\subsection{CVMFS to distribute software on grid resources}\label{section:11}

At the beginning of 2021, CVMFS was managing about 1 billion files delivered to more than 100,000 computing nodes by (i) 10 public data mirror servers - called \emph{Stratum1}s - located in Europe, Asia and the United States and (ii) 400 site-local cache servers \cite{CVMFS_2021}.

\begin{figure}[ht]
    \centering
    \includegraphics[width=0.7\textwidth]{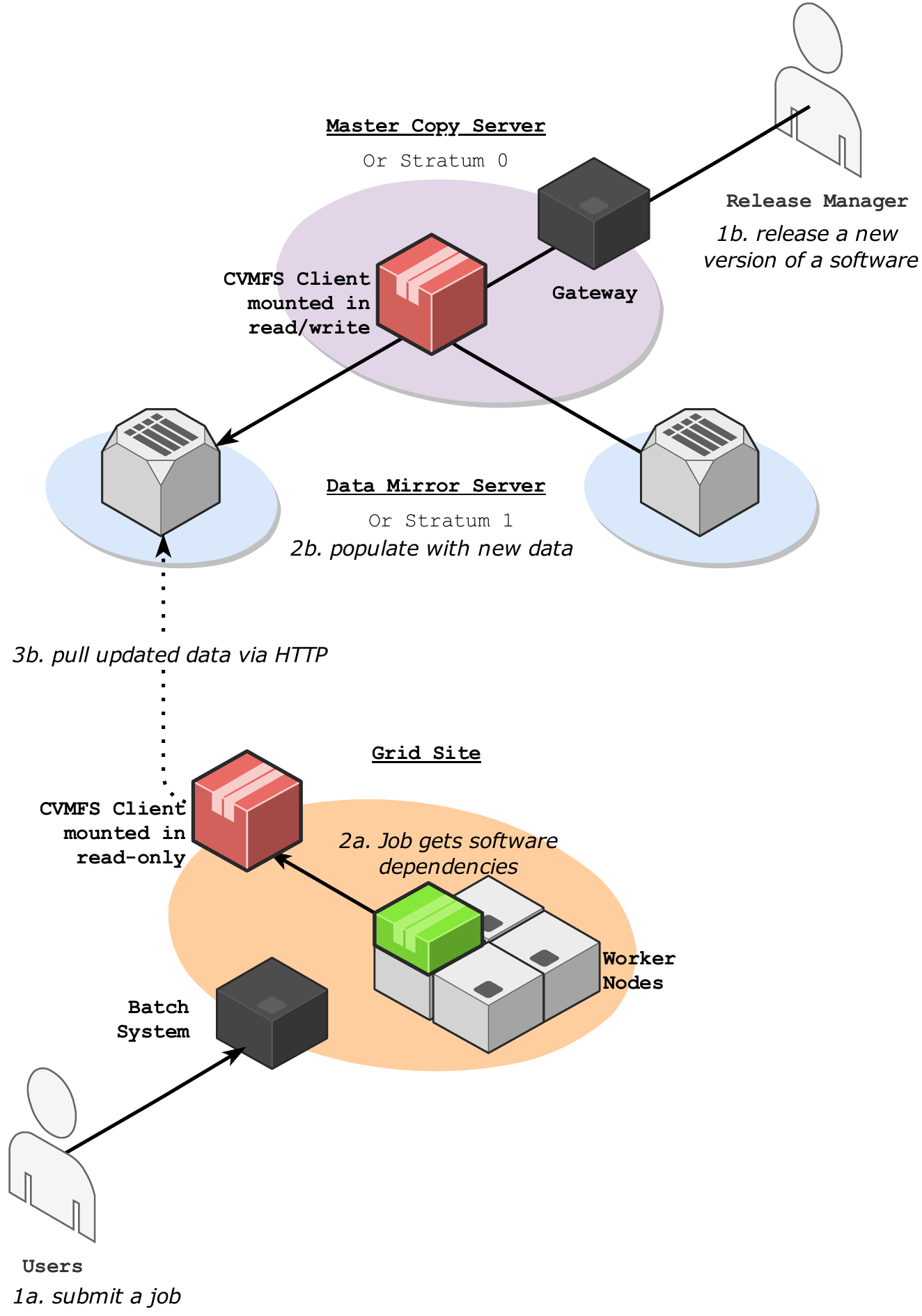}
    \caption{Schema of the CVMFS workflow on Grid Sites: (a) the steps to get software dependencies from the job; (b) the steps to publish a release of a software in CVMFS.}
    \label{fig:cvmfsGrid}
\end{figure}

To keep the file system consistent and scalable, developers conceived CVMFS as a read-only file system.
Release managers - or continuous integration workers - aiming to publish a software release has to log in to a dedicated machine - named \emph{Stratum0} - with an attached storage volume providing an authoritative and editable copy of a given repository \cite{Blomer_2019}.
Changes are written into a staging area until they are committed as a consistent changeset: new and modified files are transformed into a content-addressed object providing file-based deduplication and versioning.
In 2019, Popescu et al. \cite{Popescu_2019} introduced a gateway component, a web service in front of the authoritative storage, allowing release managers to perform concurrent operations on the same repository and make CVMFS more responsive (Figure \ref{fig:cvmfsGrid}.1.b and \ref{fig:cvmfsGrid}.2.b).

The transfer of files is then done lazily via HTTP connections initiated by the CVMFS clients \cite{Popescu_2019} (Figure \ref{fig:cvmfsGrid}.3.b).
Clients request updates based on their Time-to-Live (TTL) value, which is generally about a few minutes. Once the TTL value expires, clients download the latest version of a manifest - a text file located in the top-level directory of a given repository composed of the current root hash, metadata and the revision number of this repository - and make the updated content available.
Dykstra et al. \cite{Dykstra_2014} provide additional details about data integrity and authenticity mechanisms of CVMFS to ensure that data received matches data initially sent by a trusted server.
This pull-based approach has been proven to be robust and efficient, according to Popescu et al. \cite{Popescu_2019}, and has been widely used to distribute up-to-date software on grid sites for many years (Figure \ref{fig:cvmfsGrid}.2.a). Figure \ref{fig:cvmfsGrid} presents a simplified schema summarizing the software distribution process on grid sites via CVMFS.

Users may need to use various versions of software on heterogeneous computing resources implying different OS and architectures.
To provide a convenient environment for the users, release managers generally provide software along with build files related to many architectures, OS and compilers.
Framework for building and installing scientific software on heterogeneous systems can be used to supply CVMFS with build files.
Easybuild \cite{easybuild}, Spack \cite{spack}, Nix \cite{nix} or Gentoo \cite{gentoo} are popular choices in this area \cite{Vokl_2021,Benda_2020,Burr_2019}. 

\subsection{Software delivery on supercomputers}\label{section:12}

Communities working around the Large Hadron Collider (LHC) \cite{LHC} have extensively used WLCG and CVMFS to process a growing amount of data.
This approach was reliable during LHC Run1 but has demonstrated its limit.
According to the analysis of Stagni et al. \cite{Stagni_McNab_Luzzi_Krzemien_Consortium_2017} on the use of CPU cycles in 2016, all the LHC experiments have consumed more CPU-hours than those officially pledged to them by the WLCG: they found ways to exploit opportunistic and not officially supported resources.
Moreover, in the High-Luminosity Large Hadron Collider (HL-LHC) \cite{osti_1365580} era, experiments are expected to produce up to an order of magnitude more data compared to the current phase (LHC Run2).
To keep up with the computing needs, experiments have started to use supercomputers. 
They offer a significant amount of computing power and would potentially offer a more cost-effective data processing infrastructure compared to dedicated resources in the form of commodity clusters, as Sciacca emphasizes \cite{Sciacca_Weber_2020}.
Nevertheless, supercomputers have more restrictive security policies than Grid Sites: they do not allow CVMFS to be mounted on the nodes by default and many of them have limited or even no external connectivity.
The LHC communities have developed different solutions and strategies to cope with the lack of CVMFS, which is a critical component to run their workflows.

Stagni et al. \cite{Stagni_Valassi_Romanovskiy_2020} rely on a close collaboration with some supercomputer centers - Cineca in Italy and CSCS in Switzerland -  to get CVMFS mounted on the worker nodes.
Nevertheless, their strategy is limited to a few supercomputers and their approach would be difficult to reproduce on a large number of supercomputers: most of them do not allow such collaboration.

To deal with the lack of CVMFS on supercomputers with outbound connectivity, Filipčič et al. studied two solutions: \emph{rsync} and \emph{Parrot}  \cite{Filipcic_Haug_Hostettler_Walker_Weber_2015}.
The first solution consisted in copying the CVMFS software repository in the shared file system using \emph{rsync}: a utility aiming to transfer and synchronize files and directories between two different systems.
\emph{rsync} added a significant load on the shared file system of the supercomputers and required changes in the repository absolute paths.
The second solution was based on Parrot: a utility copied on the shared file system of the supercomputer, usable without any user privileges. 
Parrot is a wrapper using \emph{ptrace} attached to a process that intercepts system calls that access the file system and can simulate the presence of arbitrary file system mounts, CVMFS in this case.
Nevertheless, the solution was "unreliable in a multi-threaded environment" \cite{Filipcic_Haug_Hostettler_Walker_Weber_2015} because it was unable to handle race conditions. 
These methods did not constitute a production-level solution but contributed to further and future advanced solutions.

In recent years, developments in the Fuse user space libraries and the Linux kernel have lifted restrictions for mounting Fuse file systems such as CVMFS.
Developers of CVMFS have integrated these changes and designed a package called \emph{cvmfsexec} \cite{cvmfsexec}, which allows mounting the file system as an unprivileged user. 
The program needs a specific environment to work correctly: (i) external connectivity; (ii) the \emph{fusermount} library or unprivileged namespace mount points or a setuid installation of \emph{Singularity} (efficient High-Performance Computing container technology).
Blomer et al. provide additional details about the package \cite{Blomer_2020}.

Communities exploiting supercomputers that do not provide outbound connectivity cannot directly benefit from \emph{cvmfsexec}: the package still needs to pull updated data via HTTP, which is not available in such context.
We can distinguish two cases: (i) supercomputers that grant outside network or specific service access to a limited number of nodes and (ii) supercomputers that do not provide nodes with any external connectivity at all.

Tovar et al. recently worked on the first case \cite{Tovar_2021}.
They managed to build a virtual private network (VPN) client and server to redirect network traffic from the workloads running on the worker nodes to external services such as CVMFS.
In this configuration, the VPN client runs on a worker node along with the job, while the VPN server is hosted on one of the specific nodes of the supercomputer and can interact with external services.
Communities working on supercomputers from the second case cannot leverage the solution developed by Tovar et al.

O'Brien et al., one of the first teams to work with supercomputers in the LHC context, address the lack of external network access by copying part of it to the shared Lustre file system accessible by the WNs \cite{OBrien_Walker_Washbrook_2014}.
The approach (i) worked because the environment of the supercomputer was similar to a grid site one, (ii) required changes in the CVMFS files and (iii) degraded the performance of the software as Angius et al. described \cite{Angius_Oleynik_Panitkin_Turilli_De_Klimentov_Oral_Wells_Jha_2017}. 
To tackle the latter issue on the Titan supercomputer, Angius et al. moved the software to a read-only NFS server \cite{Angius_Oleynik_Panitkin_Turilli_De_Klimentov_Oral_Wells_Jha_2017}: this eliminated the problem of metadata contention and improved metadata read performance.

Similarly, on the Chinese HPC CNGrid, Filipčič regularly packed a part of CVMFS in a tarball.
Filipčič provided a deployment script to install the software and fix the path relocation on the shared file system to the local system administrators: they were then responsible for getting and updating the CVMFS tarball on the network when requested \cite{Filipcic_2017}.

To help communities to unpack a CVMFS repository in a file system, a team of developers designed \emph{uncvmfs} \cite{uncvmfs}. The utility deduplicates files of a software stack: it populates a given directory with the CVMFS files that are then hard-linked into it, if possible.
The program was used, in combination with Shifter \cite{Gerhardt_Bhimji_Canon_Fasel_Jacobsen_Mustafa_Porter_Tsulaia_2017}, a container technology providing a reproducible environment, in the context of the integration of the ALICE and CMS experiments workflows on the NERSC High-Performance Computing resources \cite{Fasel_2016,Hufnagel_2017}.
As a proof of concept, Gerhardt et al. used \emph{uncvmfs} to deduplicate the ATLAS repository and copy it into an ext4 image  - about 3.5Tb of data containing 50 million files and directories -, compressed into a 300Gb squashfs image; and Shifter to provide a software-compatible environment to run the jobs \cite{Gerhardt_Bhimji_Canon_Fasel_Jacobsen_Mustafa_Porter_Tsulaia_2017}.
Despite encapsulating the files in a container reduced the startup time of the applications, the solution generated large images, long to update and deliver on time.

To cope with large images, Teuber and the CVMFS developers conceived \emph{cvmfs\_shrinkwrap} \cite{Teuber_2019}.
The tool supports \emph{uncvmfs} features with certain optimizations and delivers additional features: \emph{cvmfs\_shrinkwrap} can extract specific files and directories based on specification files, deduplicate them, making them easy to export in various formats such as squashfs or tarball.
In this way, the following operations remain on behalf of the user communities: (i) trace their applications - meaning, in this context,  "capturing all their dependencies and their locations in the file system" -, (ii) call \emph{cvmfs\_shrinkwrap} to get a subset of CVMFS composed of the minimum required files, and (iii) export this subset in a certain format and deploy it on sequestered computing resources to run their jobs.

Douglas et al. already described such a project in an article \cite{Douglas_2019}, but the work remains specific to the ATLAS experiment.
They use \emph{uncvmfs} to produce a large image that has to be filtered afterward.
In this paper, we aim at assisting various user communities in this process by providing an open-source utility that would take applications of interest in input and would output - with the help of \emph{cvmfs\_shrinkwrap} - a subset of CVMFS with the minimum required files to run the given applications, in combination with a container image if needed.
To our knowledge, no paper has already covered the subject.

\section{Design of the CVMFS Subset Builder}\label{section:2}
\subsection{Input and output data}\label{section:21}

\begin{figure}[ht]
    \centering
    \input{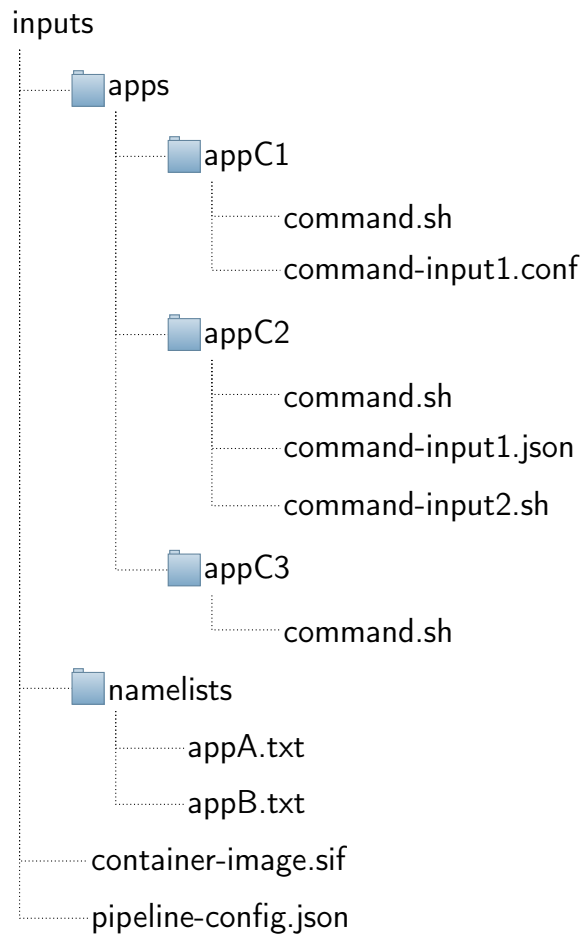}
    \caption{Schema of the input structure given to the utility.}
    \label{fig:inputTree}
\end{figure}

The utility takes a directory as input that should contain: (i) a list of applications of interest (\texttt{apps}): a command along with its input data in a separate sub-directory for each application to trace; and/or (ii) a list of files composed of paths to include in the subset of CVMFS (\texttt{namelists}).
Additionally, user communities can embed a (iii) container image compatible with Singularity to get a specific environment to trace and test the applications; (iv) and a configuration file to fine-tune the utility with variables related to the deployment process, or information about repositories.
A schema of the inputs is available in Figure \ref{fig:inputTree}.

The expected output can take different forms depending on the utility configuration:
\begin{itemize}
    \item The subset of CVMFS, generated as a standalone.
    In this case, administrators representing their user communities need to provide the right environment by themselves, which might also involve discussions with the system administrators.
    \item The subset of CVMFS embedded within the given Singularity container image.
    The utility merges both elements and submits the resulting image, which can be long to generate and deploy but may limit manual operations on the remote location. 
\end{itemize}

\subsection{Features}\label{section:22}

\begin{figure}[ht]
    \centering
    \includegraphics[width=0.74\textwidth]{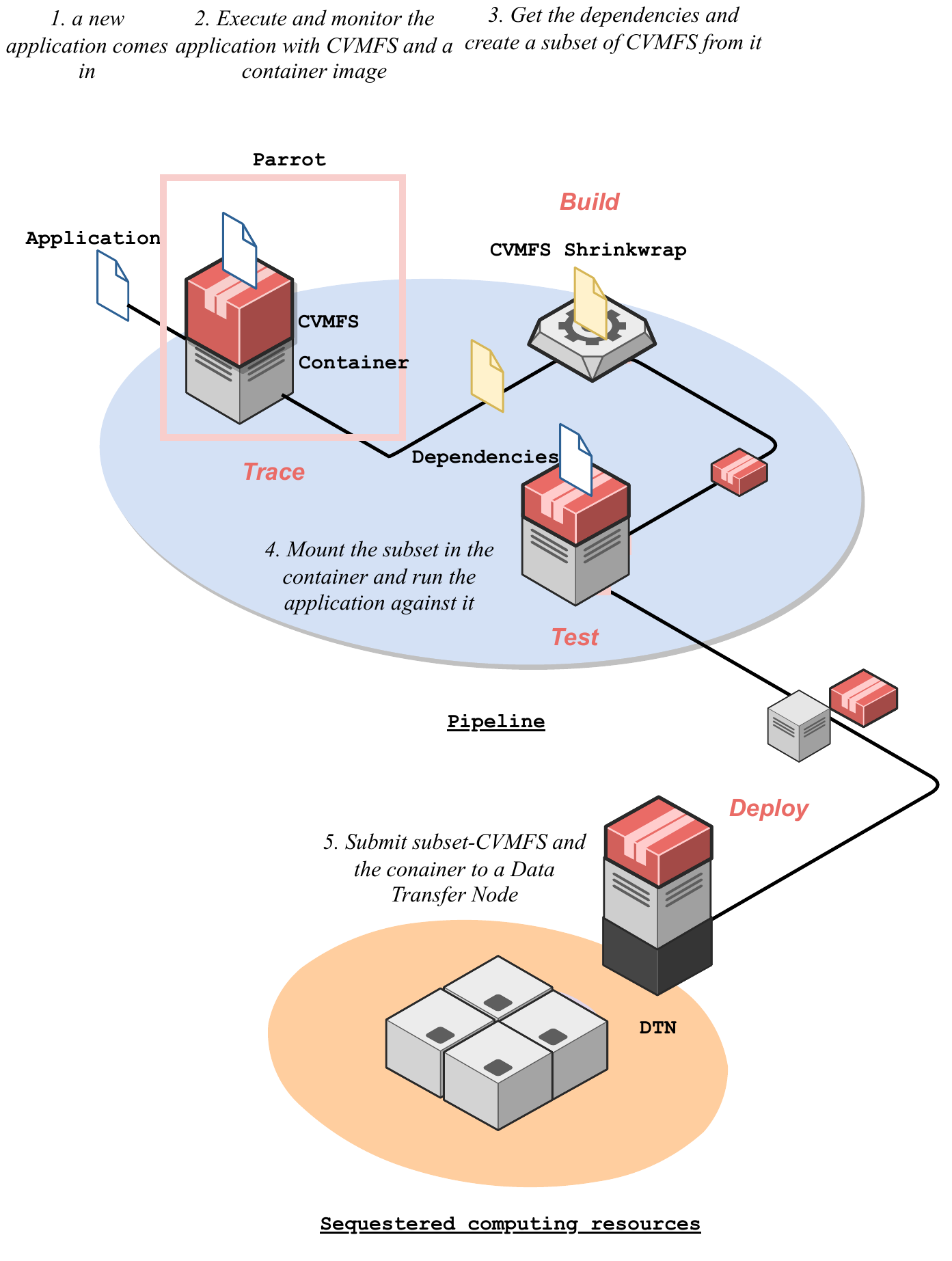}
    \caption{Schema of the utility workflow: from getting an application to trace to a subset of CVMFS on the Data Transfer Node of a High-Performance Computing cluster.}
    \label{fig:pipeline}
\end{figure}

We break down the process into four main steps, namely:
\begin{itemize}
    \item \emph{Trace}: consists in running applications contained in \texttt{apps} and trapping their system calls at runtime, using \emph{Parrot}, to identify and extract the paths of their dependencies.
    Applications can run in a Singularity container when provided, which delivers further software dependencies and a reproducible environment. 
    Dependencies are then saved in a specific file \texttt{namelist.txt}. 
    In this context, \emph{Parrot} is only used to capture system calls and, thus, is not impacted by the issues mentioned in section \ref{section:12}.
    If the step detects an error during the execution of an application, then the program is stopped.
    The step is particularly helpful for users of the utility having no technical knowledge of the applications of interest.
    \item \emph{Build}: builds a subset of CVMFS based on the paths coming from \emph{Trace} and the \texttt{namelists} directory. First, the step merges the namelist files to remove duplicated or non-existent path references, and then separates the paths in different specification files related to repositories. Finally, the step calls \emph{cvmfs\_shrinkwrap} to generate the subset of CVMFS.
    Figures \ref{fig:traceProcess} and \ref{fig:pipeline}.3 illustrate an example.
    The utility deduplicates the files, and hard-link data to populate a directory, ready to be exported in various formats as explained in Section \ref{section:12} and shown in Figure \ref{fig:pipeline}.3.

    \begin{figure}[ht]
    \centering
    \begin{minted}{yaml}
    in namelist1.txt: 
    /cvmfs/repoA/path/to/file
    /cvmfs/repoB/path/to/another/file
    in namelist2.txt:
    /cvmfs/repoA/path/to/file
    /cvmfs/repoB/path/to/yet/another/file
    
    in repoA.spec:
    /path/to/file
    in repoB.spec:
    /path/to/another/file
    /path/to/yet/another/file
    \end{minted}
    \caption{Transformation process occurring during the \emph{Trace} step: CVMFS dependencies are extracted  from \texttt{namelist.txt} and moved to specification files.}
    \label{fig:traceProcess}
    \end{figure}

    \item \emph{Test}: consists in testing certain applications - in the given Singularity container environment when provided - using the subset of CVMFS obtained during the \emph{Build} step (see Figure \ref{fig:pipeline}.4). 
    By default, applications from \texttt{apps} are used but further tests can also be provided by modifying the utility configuration.
    All the applications have to complete their execution to go to the next step.
    \item \emph{Deploy}: deploys the subset of CVMFS (Figure \ref{fig:pipeline}.5) embedded or not within the container image depending on the configuration options. If such is the case, then the utility (i) generates a new container definition file that includes the files with the container image, (ii) executes it to produce a new read-only container image. The utility supports ssh deployment via \emph{rsync}, provided the right credentials in the configuration. 
\end{itemize}

\subsection{Implementation}\label{section:23}

The utility is built as a 2-layer system.
The first layer, \emph{subcvmfs-builder}\cite{subcvmfs-builder}, is the core of the system and is self-contained.
It takes the form of a Python package, which embeds the steps described in section \ref{section:22}, and provides a command-line interface to call and execute steps independently from each other.
The first layer is, and should remain, simple and generic to be easily managed by developers and used by various communities. 

The second layer is the glue code: it consists of a workflow executing - all, or some of - the steps of the first layer.
It contains the complexity required to generate and deliver a subset of dependencies according to the needs of its users.
Unlike the first layer, the second one can take several forms and each community can tailor it for its software stack.

\begin{figure}[ht]
    \centering
    \includegraphics[width=0.8\textwidth]{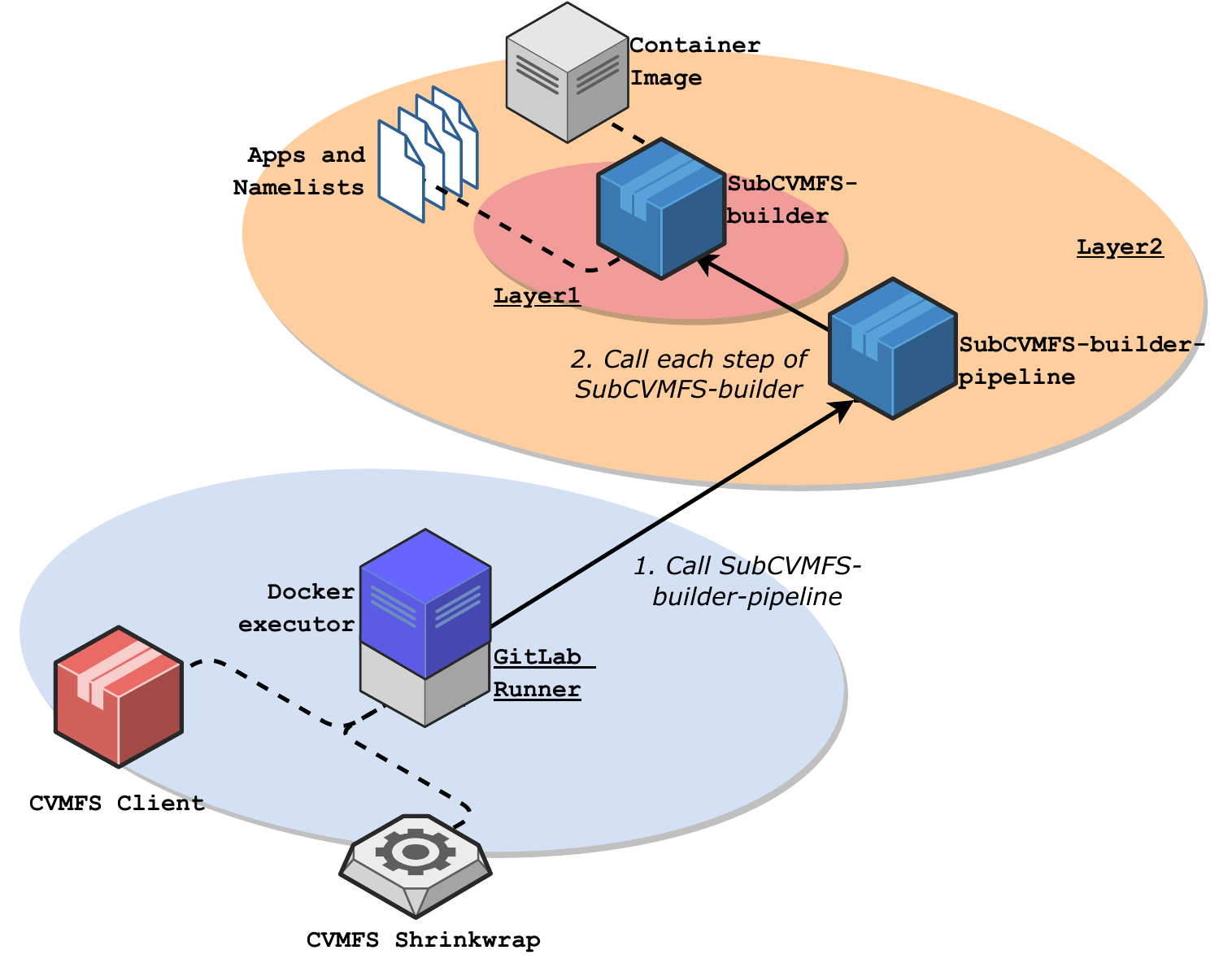}
    \caption{Schema of a layer-2 implementation within GitLab CI.}
    \label{fig:subcvmfsbuilderpipeline}
\end{figure}

We propose a first, simple and generic layer-2 implementation calling each step one after the other: \emph{subcvmfs-builder-pipeline}\cite{subcvmfs-builder-pipeline}.
This layer-2 implementation is executed from a GitLab CI/CD \cite{gitlabci}, which provides a runner and a docker executor bound to a CVMFS client to execute the code (see Figure \ref{fig:subcvmfsbuilderpipeline}) 
GitLab includes features such as log preservation to help debug the implementation and integrates a pipeline scheduling mechanism to regularly update a subset of dependencies. 
Even though this layer-2 solution is adapted for basic examples - implying a few commands to trace and test, having a small number of dependencies -, it might require further fine-tuning for more advanced use cases.

Indeed, this generic layer-2 implementation is not scalable as it (i) is a single-threaded and single-process program, and (ii) requires manual operations to insert additional inputs in the process.
This is not adapted to communities having to trace and test hundreds of various applications to generate large subsets of CVMFS.
Two possibilities for such communities: building a new layer-2 implementation - able to automatically fetch applications and trace/test them in parallel - based on \emph{subcvmfs-builder-pipeline} or creating one from scratch.

In the next section, we are going to study how the LHCb experiment \cite{LHCb_2008} leverages \emph{subcvmfs-builder} and \emph{subcvmfs-builder-pipeline} to deliver Gauss \cite{Clemencic_2011}, a Monte-Carlo simulation program, on the worker nodes of Mare Nostrum \cite{David_2010}, a supercomputer with no external connectivity based in Barcelona, Spain.

\section{A Practical Use Case}\label{section:3}
\subsection{Gauss}\label{section:31}

To better understand experimental conditions and performances, the LHCb collaboration has developed Gauss, a Monte-Carlo simulation application - based on the Gaudi framework \cite{Barrand_2001} - that reproduces events occurring in the LHCb detector.
The application consists of two independent phases executed sequentially, namely the generation of the events \cite{Belyaev_2011} relying on Pythia \cite{Sj_strand_2001} by default; the tracking of the particles through the simulated detector depending on Geant4 \cite{geant_2003}.

In 2021, Gauss represents about 70\% of the distributed computing activities of the LHCb collaboration and 150 million events are simulated per day.
The application has originally been tailored for WLCG grid sites: Gauss is a compute-intensive single-process (SP), single-threaded (ST) application, only supporting x86 architectures and CERN-CentOS-compatible environments \cite{LinuxWebCERN}.
Gauss and most of its dependencies are delivered via CVMFS.

Gauss takes a certain number of events to process as inputs, as well as a "run number" and an "event number".
The combination of both numbers forms a seed, which ensures repeatability during the generation and simulation phases.
It mainly relies on packages such as Python, Boost and gcc to produce histograms and \emph{ntuples} under the form of a ROOT \cite{root} file.

\begin{figure}[ht]
    \centering
    \includegraphics[width=0.8\textwidth]{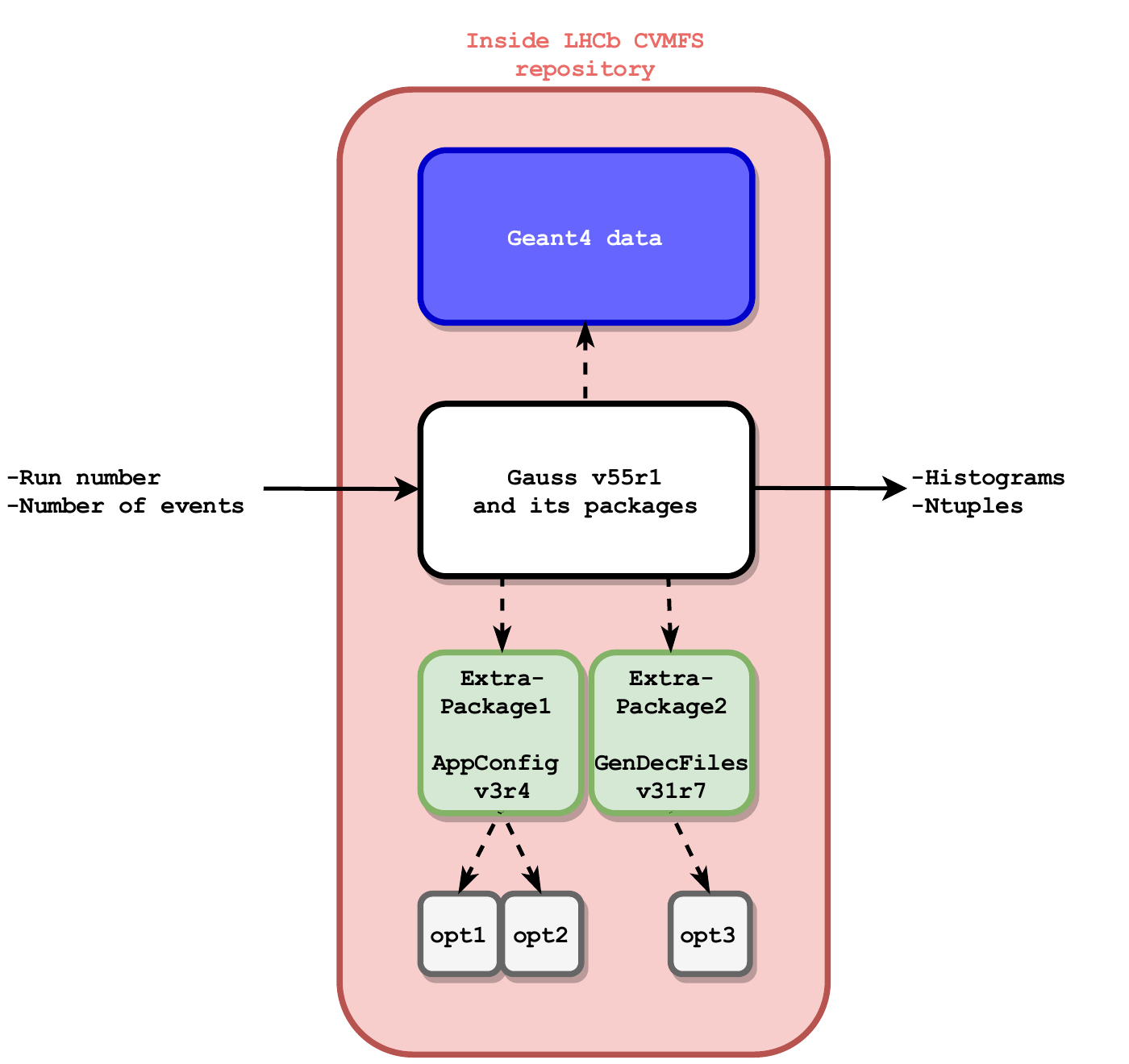}
    \caption{Example of a Gauss instance, its dependencies and some interactions with extra packages and their options.}
    \label{fig:gauss}
\end{figure}

Gauss is modular and highly configurable and constitutes a complex use-case: it can integrate extra packages such as various event generators and decay tools. 
Depending on LHCb production needs and the computing environments available, different versions of Gauss and its attached packages can be used.
A plethora of option files can also be passed as inputs to the extra packages.
Figure \ref{fig:gauss} describes the inputs, outputs and dependencies of Gauss as well as its interactions with some extra packages and their options.

\subsection{Mare Nostrum}\label{section:32}

To start integrating their workflows on High-Performance computing resources, LHC experiments can benefit from a collaboration with PRACE \cite{prace} and GÉANT \cite{geant,collabhpc}.
This collaboration gives them access to several European supercomputers such as Marconi in Italy and Mare Nostrum in Spain.

Managed by the Barcelona Supercomputing Center (BSC), MareNostrum is the most powerful and emblematic supercomputer in Spain \cite{marenostrum}.
MareNostrum was built in 2004 (MareNostrum 1), has been updated 3 times since then (Mare Nostrum 2,3 and 4) and was ranked 63rd in the June 2021 Top500 list \cite{top500}.
Each node composing the general-purpose block is equipped with two Intel Xeon Platinum 8160 24 cores at 2.1 GHz chips, and at least 2GB of RAM: this configuration matches with Gauss requirements.
Nevertheless, Mare Nostrum is more restrictive than a traditional Grid Site on WLCG: (i) no external connectivity at all; (ii) no service can be installed on the edge node; (iii) no CVMFS, and thus, no Gauss and its dependencies available.


\subsection{Running Gauss on Mare Nostrum}\label{section:33}

Running embarrassingly parallel applications such as Gauss on a supercomputer can be seen as counterproductive.
While it is true that the interconnect of the supercomputer partitions has not been designed for millions of small Monte-Carlo runs, it is better to use available, otherwise unused, cycles in agreement with the management of the supercomputer sites.
In the meantime, developers are adapting software \cite{Siddi_Muller_2019,Mazurek_2021}, but it remains a long process, requiring deep and technical software inputs.

To deliver Gauss on Mare Nostrum, LHCb can rely on (i) \emph{subcvmfs-builder} to produce a subset of CVMFS containing the required files; (ii) a CernVM Singularity container to provide a Gauss-compatible environment and to mount the subset of CVMFS as if it was a CVMFS client.

Nevertheless, as we explained in \ref{section:31}, a Gauss execution can involve different packages, extra packages, options, data and versions. Encapsulating its ecosystem requires a good understanding of the application and/or a large amount of storage to encapsulate the right dependencies.
Therefore, different options are available:
\begin{itemize}
    \item Include the whole LHCb CVMFS repository: would not require any specific knowledge about Gauss and would involve all the necessary files to run any Gauss instance. However, this option would imply a tremendous quantity of storage - the full LHCb repository needs 5.2 Terabytes -, long periods to update the subset and many unnecessary files.
    \item Include the dependencies of various Gauss runs: as the first option, would not need any specific knowledge about Gauss and would include a few gigabytes of data. Nevertheless, such an option would not guarantee the presence of all needed files and would require a tremendous amount of computing resources to trace Gauss workloads continuously.
    \item Include all the known dependencies of Gauss: would require a deep understanding of Gauss and its dependencies to include all the required files in a subset of CVMFS. While this option would not involve many computing or storage resources, it would include human resources to update the content of the subset of CVMFS according to the releases of Gauss and its extra packages. 
\end{itemize}

As the default storage quota on Mare Nostrum is smaller than the LHCb repository, we decided to reject the first option.
LHCb has access to tremendous computing power: it interacts with hundreds of WLCG Sites to run Gauss workloads and could theoretically trace them and extract their requirements.
In practice, tracing Gauss workloads in production could slow down the applications and their execution, which is not an option.
Similarly, LHCb does not have human resources to update the subset of CVMFS according to the changes done.
Thus, we chose to combine the second and the third options to propose a light and easy to update and maintain solution.
The process consists in getting insights into the structure of the Gauss dependencies by running and tracing a small set of Gauss workloads and analyzing the system calls before including the structure in \emph{subcvmfs-builder-pipeline}.

After analyzing 500 commands calling Gauss from the LHCb production environment and tracing 3 Gauss applications using \emph{subcvmfs-builder} \cite{gauss_analysis}, we noticed that:
\begin{itemize}
    \item 97\% of the workloads studied were running the same Gauss versions (v49r20) with the same extra packages and versions.
    The versions of Gauss and its extra packages seem related to the underlying architecture.
    \item 846 Mb of files were needed to run 3 Gauss (v49r20) workloads.
    About 95\% of the size is related to the Gauss version and the underlying architecture, and is common to the Gauss workloads traced, while the 5\% left is bound to the options and Geant4 data used that are specific to a given Gauss workload.
    \item Integrating all the options and Geant4 data related to Gauss v49r20 would correspond to 1.8 Gb of files.
\end{itemize}

Based on these assumptions, we created a \texttt{namelist} file containing (i) the files shared by the 3 Gauss applications that we traced and (ii) all the options and Geant4 data in order to generate a subset of CVMFS able to run any Gauss workload targeting the v49r20 version.
We used \emph{subcvmfs-builder-pipeline} to build the subset of CVMFS, to successfully test it with 5 Gauss workloads - different from the ones we used previously - and to deploy it to Mare Nostrum.
We fine-tuned the utility to disable the \emph{trace} step and to deploy the subset separately from the container.
Indeed, CernVM - the container that we use to provide a reproducible environment to the workload - does not need regular updates and merging it with the subset of CVMFS is a time-consuming operation.

This resulted in a CernVM singularity container occupying 6.4 Gb on the General Parallel File System (GPFS) of Mare Nostrum combined with a subset of CVMFS covering 6 Gb: dependencies occupies 3.2 Gb of space while 2.8 Gb are required for the \emph{cvmfs\_shrinkwrap} metadata.
Thus, 12.4 Gb of space on the GPFS of Mare Nostrum is currently sufficient to run 97\% of the Gauss workloads analyzed: 0.24\% of the LHCb repository.

Even though this approach provides a light, easy and fast-to-update solution, LHCb developers need to keep it up to date to integrate new versions or structure changes.
One way to proceed would consist in automating and repeating the analysis work regularly. 
One could also integrate the \emph{trace} command of \emph{subcvmfs-builder} within the LHCb production test phase, which consists in running a few events of upcoming Gauss workloads on a given Grid Site.
LHCb developers could trace some of them during the process and store the traces in a database.
An LHCb-specific \emph{subcvmfs-pipeline-builder} could then periodically fetch the content of the database to build, test and deploy a new subset of dependencies to Mare Nostrum.


\section{Conclusion}
This paper presents a dependency delivery system based on CVMFS to provide complex software stacks on sequestered computing resources such as worker nodes of supercomputers not having external connectivity.

After introducing CVMFS (section \ref{section:11}), a critical tool - especially for LHC communities - to supply workloads with complex dependencies on Grid Sites, we have described the context of this study (section \ref{section:12}): several virtual organizations are exporting their workflow from WLCG to supercomputers, which have more restrictive policies than grid sites and generally do not allow to mount CVMFS on the worker nodes.

We have highlighted several solutions aiming to overcome the issue such as collaborating with the system administrators and using tools such as \emph{Parrot} and \emph{cvmfsexec}.
Nevertheless, these approaches do not work when worker nodes have no external connectivity.
Then, we have emphasized different ways to export parts of CVMFS to supercomputers with no external connectivity: \emph{uncvmfs} and \emph{cvmfs\_shrinkwrap}.
These solutions require several manual steps and therefore we have proposed a utility to assist communities in this process.

We have explained the different steps of the utility in detail (section \ref{section:22}). It traces - captures the system calls of - applications of interest, builds a subset with the required files, tests the subset and deploys it to a remote computing resource.
We also described the structure of the solution (section \ref{section:23}), which is composed of two layers: a first one, generic with simple components, and a second one more complex, adapted to communities needs that can be fine-tuned.

Finally, we have provided a use case based on Gauss, a Monte-Carlo simulation application reproducing events occurring in the LHCb detector (section \ref{section:31}).
Gauss is highly configurable and can be coupled with different packages, extra packages, options, data and versions.
It represents a complex bundle of dependencies, which makes it ideal to test our utility.
We have proposed a method to encapsulate Gauss and its dependencies in a subset, which represents 12.4 Gb of space on the GPFS of the Mare Nostrum supercomputer (section \ref{section:33}).
The solution produced represents 0.24\% of the full LHCb repository and, thus, is easier to update.
We have successfully tested the solution with different Gauss workloads.
Future work could focus on encapsulating further applications from different domains using this utility, and analyzing its performances to deploy subsets on various supercomputers.

%
%
%
\bibliographystyle{plain}
\bibliography{summary}

\begin{thebibliography}{10}

\bibitem{geant_2003}
S.~Agostinelli and al.
\newblock Geant4—a simulation toolkit.
\newblock {\em Nuclear Instruments and Methods in Physics Research Section A:
  Accelerators, Spectrometers, Detectors and Associated Equipment}, 506(3):250
  -- 303, 2003.

\bibitem{osti_1365580}
G.~Apollinari, I.~Béjar~Alonso, O.~Brüning, M.~Lamont, and L.~Rossi.
\newblock {\em High-Luminosity Large Hadron Collider (HL-LHC) : Preliminary
  Design Report}.
\newblock CERN Yellow Reports: Monographs. CERN, Geneva, Dec. 2015.

\bibitem{Arsuaga_R_os_2015}
Mar{\'{\i}}a Arsuaga-R{\'{\i}}os, Seppo~S Heikkilä, Dirk Duellmann, Ren{\'{e}}
  Meusel, Jakob Blomer, and Ben Couturier.
\newblock Using s3 cloud storage with {ROOT} and {CvmFS}.
\newblock {\em Journal of Physics: Conference Series}, 664(2):022001, dec 2015.

\bibitem{Barrand_2001}
G.~Barrand, I.~Belyaev, P.~Binko, M.~Cattaneo, R.~Chytracek, G.~Corti,
  M.~Frank, G.~Gracia, J.~Harvey, E.van Herwijnen, P.~Maley, P.~Mato,
  S.~Probst, and F.~Ranjard.
\newblock Gaudi — a software architecture and framework for building hep data
  processing applications.
\newblock {\em Computer Physics Communications}, 140(1):45--55, 2001.
\newblock CHEP2000.

\bibitem{Barreiro_2019}
Fernando Barreiro, Doug Benjamin, Taylor Childers, Kaushik De, Johannes
  Elmsheuser, Andrej Filipčič, Alexei Klimentov, Mario Lassnig, Tadashi
  Maeno, Danila Oleynik, and et~al.
\newblock The future of distributed computing systems in atlas: Boldly
  venturing beyond grids.
\newblock {\em EPJ Web of Conferences}, 214:03047, 2019.

\bibitem{Belyaev_2011}
I~Belyaev, T~Brambach, N~H Brook, N~Gauvin, G~Corti, K~Harrison, P~F Harrison,
  J~He, C~R Jones, M~Lieng, G~Manca, S~Miglioranzi, P~Robbe, V~Vagnoni,
  M~Whitehead, and J~Wishahi and.
\newblock Handling of the generation of primary events in gauss, the {LHCb}
  simulation framework.
\newblock {\em Journal of Physics: Conference Series}, 331(3):032047, Dec.
  2011.

\bibitem{Douglas_2019}
{Benjamin, Douglas}, {Childers, Taylor}, {Lesny, David}, {Oleynik, Danila},
  {Panitkin, Sergey}, {Tsulaia, Vakho}, {Yang, Wei}, and {Zhao, Xin}.
\newblock Building and using containers at hpc centres for the atlas
  experiment.
\newblock {\em EPJ Web Conf.}, 214:07005, 2019.

\bibitem{CVMFS_2021}
Jakob Blomer.
\newblock Cernvm-fs overview and roadmap.
\newblock [Online] Available: https://easybuild.io/eum/002\_eum21\_cvmfs.pdf,
  2021.
\newblock [Accessed: 26-May-2021].

\bibitem{Blomer_2017}
Jakob Blomer, Gerardo Ganis, Nikola Hardi, and Radu Popescu.
\newblock Delivering lhc software to hpc compute elements with cernvm-fs.
\newblock In Julian~M. Kunkel, Rio Yokota, Michela Taufer, and John Shalf,
  editors, {\em High Performance Computing}, pages 724--730, Cham, 2017.
  Springer International Publishing.

\bibitem{Blomer_2020}
{Blomer, Jakob}, {Dykstra, Dave}, {Ganis, Gerardo}, {Mosciatti, Simone}, and
  {Priessnitz, Jan}.
\newblock A fully unprivileged cernvm-fs.
\newblock {\em EPJ Web Conf.}, 245:07012, 2020.

\bibitem{Blomer_2019}
{Blomer, Jakob}, {Ganis, Gerardo}, {Mosciatti, Simone}, and {Popescu, Radu}.
\newblock Towards a serverless cernvm-fs.
\newblock {\em EPJ Web Conf.}, 214:09007, 2019.

\bibitem{subcvmfs-builder}
Alexandre~Franck Boyer.
\newblock Subcvmfs-builder.
\newblock Mar 2022.

\bibitem{subcvmfs-builder-pipeline}
Alexandre~Franck Boyer.
\newblock Subcvmfs-builder-pipeline.
\newblock Mar 2022.

\bibitem{gauss_analysis}
Alexandre~Franck Boyer.
\newblock Subcvmfs: Gauss analysis, Mar 2022.

\bibitem{marenostrum}
BSC.
\newblock Marenostrum.
\newblock [Online] Available: https://www.bsc.es/marenostrum/, 2020.
\newblock [Accessed: 04-Oct-2021].

\bibitem{Buncic_2010}
P~Buncic, C~Aguado Sanchez, J~Blomer, L~Franco, A~Harutyunian, P~Mato, and
  Y~Yao.
\newblock {CernVM} {\textendash} a virtual software appliance for {LHC}
  applications.
\newblock {\em Journal of Physics: Conference Series}, 219(4):042003, apr 2010.

\bibitem{Burr_2019}
{Burr, Chris}, {Clemencic, Marco}, and {Couturier, Ben}.
\newblock Software packaging and distribution for lhcb using nix.
\newblock {\em EPJ Web Conf.}, 214:05005, 2019.

\bibitem{collabhpc}
CERN.
\newblock Cern, skao, gÉant and prace to collaborate on high-performance
  computing.
\newblock [Online] Available:
  https://home.cern/news/news/computing/cern-skao-geant-and-prace-collaborate-high-performance-computing,
  2020.
\newblock [Accessed: 04-Oct-2021].

\bibitem{LinuxWebCERN}
CERN.
\newblock Linux@cern.
\newblock [Online] Available: https://linux.web.cern.ch/, 2020.
\newblock [Accessed: 09-Feb-2021].

\bibitem{CVMFS}
CERN.
\newblock Cernvm-fs.
\newblock [Online] Available: https://cernvm.cern.ch/, 2021.
\newblock [Accessed: 19-May-2021].

\bibitem{LHC}
CERN.
\newblock The large hadron collider.
\newblock [Online] Available:
  https://home.cern/science/accelerators/large-hadron-collider, 2021.
\newblock [Accessed: 27-May-2021].

\bibitem{root}
CERN.
\newblock Root: analyzing petabytes of data, scientifically.
\newblock [Online] Available: https://root.cern.ch/, 2021.
\newblock [Accessed: 30-Sep-2021].

\bibitem{WLCG}
CERN.
\newblock Worldwide lhc computing grid.
\newblock [Online] Available: https://wlcg.web.cern.ch/, 2021.
\newblock [Accessed: 27-May-2021].

\bibitem{Clemencic_2011}
M~Clemencic, G~Corti, S~Easo, C~R Jones, S~Miglioranzi, M~Pappagallo, and
  P~Robbe and.
\newblock The {LHCb} simulation application, gauss: Design, evolution and
  experience.
\newblock {\em Journal of Physics: Conference Series}, 331(3):032023, Dec.
  2011.

\bibitem{LHCb_2008}
The~LHCb Collaboration.
\newblock The {LHCb} detector at the {LHC}.
\newblock {\em Journal of Instrumentation}, 3(08):S08005--S08005, aug 2008.

\bibitem{cvmfsexec}
CVMFS.
\newblock cvmfsexec.
\newblock [Online] Available: https://github.com/cvmfs/cvmfsexec, 2021.
\newblock [Accessed: 28-May-2021].

\bibitem{Dykstra_2014}
D.~{Dykstra} and J.~{Blomer}.
\newblock {Security in the CernVM File System and the Frontier Distributed
  Database Caching System}.
\newblock In {\em Journal of Physics Conference Series}, volume 513 of {\em
  Journal of Physics Conference Series}, page 042015, June 2014.

\bibitem{easybuild}
EasyBuild.
\newblock Easybuild: building software with ease.
\newblock [Online] Available: https://easybuild.io/, 2021.
\newblock [Accessed: 11-Dec-2021].

\bibitem{Fasel_2016}
Markus Fasel.
\newblock Using nersc high-performance computing (hpc) systems for high-energy
  nuclear physics applications with alice.
\newblock {\em Journal of Physics: Conference Series}, 762:012031, Oct. 2016.

\bibitem{Filipcic_2017}
Andrej Filipčič.
\newblock Integration of the chinese hpc grid in atlas distributed computing.
\newblock {\em Journal of Physics: Conference Series}, 898:082008, Oct. 2017.

\bibitem{Filipcic_Haug_Hostettler_Walker_Weber_2015}
Andrej Filipčič, S.~Haug, Michi Hostettler, Rodney Walker, and Michele Weber.
\newblock Atlas computing on cscs hpc.
\newblock {\em Journal of Physics: Conference Series}, 664(9):092011, Dec.
  2015.

\bibitem{gentoo}
Gentoo.
\newblock Gentoo linux.
\newblock [Online] Available: https://www.gentoo.org/, 2021.
\newblock [Accessed: 11-Dec-2021].

\bibitem{Gerhardt_Bhimji_Canon_Fasel_Jacobsen_Mustafa_Porter_Tsulaia_2017}
Lisa Gerhardt, Wahid Bhimji, Shane Canon, Markus Fasel, Doug Jacobsen, Mustafa
  Mustafa, Jeff Porter, and Vakho Tsulaia.
\newblock Shifter: Containers for hpc.
\newblock {\em Journal of Physics: Conference Series}, 898:082021, Oct. 2017.

\bibitem{gitlabci}
GitLab.
\newblock Gitlab ci/cd.
\newblock [Online] Available: https://docs.gitlab.com/ee/ci/, 2021.
\newblock [Accessed: 23-Sep-2021].

\bibitem{geant}
GÉANT.
\newblock GÉant.
\newblock [Online] Available: https://www.geant.org/, 2021.
\newblock [Accessed: 04-Oct-2021].

\bibitem{Harutyunyan_2012}
A~Harutyunyan, J~Blomer, P~Buncic, I~Charalampidis, F~Grey, A~Karneyeu,
  D~Larsen, D~Lombra{\~{n}}a Gonz{\'{a}}lez, J~Lisec, B~Segal, and P~Skands.
\newblock {CernVM} co-pilot: an extensible framework for building scalable
  computing infrastructures on the cloud.
\newblock {\em Journal of Physics: Conference Series}, 396(3):032054, dec 2012.

\bibitem{Hufnagel_2017}
Dirk Hufnagel.
\newblock Cms use of allocation based hpc resources.
\newblock {\em Journal of Physics: Conference Series}, 898:092050, Oct. 2017.

\bibitem{uncvmfs}
ic~hep.
\newblock uncvmfs.
\newblock [Online] Available: https://github.com/ic-hep/uncvmfs, 2018.
\newblock [Accessed: 30-May-2021].

\bibitem{Mazurek_2021}
Michal Mazurek, Gloria Corti, and Dominik Muller.
\newblock {New simulation software technologies at the LHCb Experiment at
  CERN}.
\newblock 12 2021.

\bibitem{nix}
NixOS.
\newblock Nixos.
\newblock [Online] Available: https://nixos.org/, 2021.
\newblock [Accessed: 11-Dec-2021].

\bibitem{Angius_Oleynik_Panitkin_Turilli_De_Klimentov_Oral_Wells_Jha_2017}
Danila Oleynik, Sergey Panitkin, Matteo Turilli, Alessio Angius, Kaushik De,
  Alexei Klimentov, Sarp~H. Oral, Jack~C. Wells, and Shantenu Jha.
\newblock High-throughput computing on high-performance platforms: A case
  study, 2017.

\bibitem{OBrien_Walker_Washbrook_2014}
B.~O’Brien, R.~Walker, and A.~Washbrook.
\newblock Leveraging hpc resources for high energy physics.
\newblock {\em Journal of Physics: Conference Series}, 513(3):032104, Jun.
  2014.

\bibitem{Popescu_2019}
{Popescu, Radu}, {Blomer, Jakob}, and {Ganis, Gerardo}.
\newblock Towards a responsive cernvm-fs architecture.
\newblock {\em EPJ Web Conf.}, 214:03036, 2019.

\bibitem{prace}
PRACE.
\newblock Partnership for advanced computing in europe.
\newblock [Online] Available: https://prace-ri.eu/, 2021.
\newblock [Accessed: 04-Oct-2021].

\bibitem{Sciacca_Weber_2020}
Francesco~Giovanni Sciacca.
\newblock Enabling atlas big data processing on piz daint at cscs.
\newblock {\em EPJ Web Conf.}, 245:09005, 2020.

\bibitem{Buncic_2011}
B~Segal, P~Buncic, C~Aguado Sanchez, J~Blomer, D~Garcia Quintas, A~Harutyunian,
  P~Mato, J~Rantala, D~Weir, and Y~Yao.
\newblock Lhc cloud computing with cernvm.
\newblock {\em 13th International Workshop on Advanced Computing and Analysis
  Techniques in Physics Research (ACAT2010)}, 093(4):042003, feb 2011.

\bibitem{Siddi_Muller_2019}
Benedetto~Gianluca Siddi and Dominik Müller.
\newblock Gaussino - a gaudi-based core simulation framework.
\newblock In {\em 2019 IEEE Nuclear Science Symposium and Medical Imaging
  Conference (NSS/MIC)}, page 1–4, Manchester, United Kingdom, Oct. 2019.
  IEEE.

\bibitem{Sj_strand_2001}
Torbjörn Sjöstrand, Patrik Edén, Christer Friberg, Leif Lönnblad, Gabriela
  Miu, Stephen Mrenna, and Emanuel Norrbin.
\newblock High-energy-physics event generation with pythia 6.1.
\newblock {\em Computer Physics Communications}, 135(2):238–259, Apr. 2001.

\bibitem{spack}
Spack.
\newblock Spack.
\newblock [Online] Available: https://spack.readthedocs.io/en/latest/, 2021.
\newblock [Accessed: 11-Dec-2021].

\bibitem{Stagni_McNab_Luzzi_Krzemien_Consortium_2017}
Federico Stagni, Andrew McNab, Cinzia Luzzi, Wojciech Krzemien, and Dirac
  Consortium.
\newblock Dirac universal pilots.
\newblock {\em Journal of Physics: Conference Series}, 898(9):092024, Oct.
  2017.

\bibitem{Stagni_Valassi_Romanovskiy_2020}
Federico Stagni, Andrea Valassi, and Vladimir Romanovskiy.
\newblock Integrating lhcb workflows on hpc resources: status and strategies.
\newblock {\em EPJ Web of Conferences}, 245:09002, 2020.

\bibitem{Teuber_2019}
Samuel Teuber.
\newblock {Efficient unpacking of required software from CERNVM-FS}, February
  2019.

\bibitem{top500}
Top500.
\newblock Top500.
\newblock [Online] Available: https://www.top500.org/, 2021.
\newblock [Accessed: 04-Oct-2021].

\bibitem{Tovar_2021}
{Tovar, Benjamin}, {Bockelman, Brian}, {Hildreth, Michael}, {Lannon, Kevin},
  and {Thain, Douglas}.
\newblock Harnessing hpc resources for cms jobs using a virtual private
  network.
\newblock {\em EPJ Web Conf.}, 251:02032, 2021.

\bibitem{David_2010}
David Vicente and Javier Bartolome.
\newblock Bsc-cns research and supercomputing resources.
\newblock In Michael Resch, Sabine Roller, Katharina Benkert, Martin Galle,
  Wolfgang Bez, and Hiroaki Kobayashi, editors, {\em High Performance Computing
  on Vector Systems 2009}, pages 23--30, Berlin, Heidelberg, 2010. Springer
  Berlin Heidelberg.

\bibitem{Vokl_2021}
{Volkl, Valentin}, {Madlener, Thomas}, {Lin, Tao}, {Wang, Joseph},
  {Konstantinov, Dmitri}, {Razumov, Ivan}, {Sailer, Andre}, and {Ganis,
  Gerardo}.
\newblock Building hep software with spack: Experiences from pilot builds for
  key4hep and outlook for lcg releases.
\newblock {\em EPJ Web Conf.}, 251:03056, 2021.

\bibitem{Benda_2020}
{Xu, Benda}, {Amadio, Guilherme}, {Gro, Fabian}, and {Haubenwallner, Michael}.
\newblock Gentoo prefix as a physics software manager.
\newblock {\em EPJ Web Conf.}, 245:05036, 2020.

\end{thebibliography}
\end{document}